# What you know or who you know? The role of intellectual and social capital in opportunity recognition

José-Aurelio Medina-Garrido, Antonio-Rafael Ramos-Rodríguez, José-Daniel Lorenzo-Gómez, and José Ruiz-Navarro
Department of Business Organization, University of Cadiz, Spain

This was the version submitted to the "International Small Business Journal" and accepted for publication. The final published version can be found at https://doi.org/10.1177/0266242610369753
We acknowledge that SAGE PUBLICATIONS LTD holds the copyright of the final version of this work.
Please, cite this paper in this way: Ramos-Rodríguez, Antonio-Rafael; Medina-Garrido, José-Aurelio; Lorenzo-Gómez, José-Daniel; Ruiz-Navarro, José (2010). What you know or who you know? The role of intellectual and social capital in opportunity recognition. International Small Business Journal. 28(6) 566–582. ISSN 0266-2426.

**Abstract**

The recognition of business opportunities is the first stage in the entrepreneurial process. The current work analyzes the effects of individuals' possession of and access to knowledge on the probability of recognizing good business opportunities in their area of residence. The authors use an eclectic theoretical framework consisting of intellectual and social capital concepts. In particular, they analyze the role of individuals' educational level, their perception that they have the right knowledge and skills to start a business, whether they own and manage a firm, their contacts with other entrepreneurs, and whether they have been business angels. The hypotheses proposed here are tested using data collected for the GEM project in Spain in 2007. The results show that individuals' access to external knowledge through the social networks in which they participate is fundamental for developing the capacity to recognize new business opportunities.

# Introduction

It is widely recognized nowadays that the firm creation phenomenon has a positive impact on the generation of both wealth and employment in the country (Reynolds et al., 2003; Acs et al., 2005). The academic literature has responded by showing more interest in studying the entrepreneurial process. In particular, researchers are increasingly interested in studying firm creation in pre-incorporation stages (Delmar and Gunnarsson, 2000; Davidsson and Honig, 2003; Wagner, 2004; Langowitz and Minniti, 2005; Arenius and Minniti, 2005). In this respect, the firm creation process begins with the recognition of opportunities. Opportunities are a fundamental requirement in entrepreneurial activities: without opportunities, firm creation is impossible.

For Kirzner (1973), entrepreneurship is the ability to perceive new opportunities. According to Stevenson, Roberts and Grousbeck (1999), entrepreneurship involves pursuing an opportunity regardless of the resources or capabilities currently available. Shane and Venkataraman (2000: 218) define the field of entrepreneurship as the study of "how, by whom, and with what effects opportunities to create future goods and services are discovered, evaluated, and exploited". Hitt et al. (2001) understand entrepreneurship as the identification and exploitation of opportunities not exploited previously.

Thus, there is a consensus about the connection between entrepreneurship and the identification and development of business opportunities (Smith and DeGregorio, 2001; Zahra and Dess, 2001; Alsos and Kaikkonen, 2004). The interest in business opportunities has generated various lines of research. Work has centred on defining opportunity as a concept (Gartner et al., 2003), the opportunity generation process (Craig and Lindsay, 2001; Shepherd and DeTienne, 2001; Shepherd and Levesque, 2002; Corbett, 2002; Linton and Walsh, 2008), the opportunity exploitation process (Samuelsson, 2001), and the importance of studying business opportunities in the entrepreneurship field (Eckhardt and Shane, 2003; Gartner et al., 2003).

But the existence of business opportunities alone is not sufficient. Someone must be able to identify them. In this line, the literature proposes a large number of definitions that consider the role of opportunity recognition in the entrepreneurial process. For Shane and Venkataraman (2000: 218), the field of entrepreneurship consequently involves "the study of sources of opportunities (…) and the set of individuals who discover, evaluate, and exploit them". Thus, entrepreneurship research examines the recognition and exploitation of new and existing opportunities, and the cognitive processes, behaviours and modes of action to exploit such opportunities (Meyer et al., 2002).

It is clear that the recognition of opportunities by individuals is a topic of interest for researchers in the entrepreneurship field. Some research efforts have centred on the role of individuals' prior knowledge (Shane, 2000) and of the social networks in which they participate (Arenius and De Clercq, 2005). These two aspects relate opportunity recognition to the possession of prior knowledge (which forms part of the individual's intellectual capital), and to the access to external knowledge (coming from social networks, or the social capital), respectively.

Researchers have studied the role of prior knowledge in opportunity recognition mainly from a qualitative approach (Shane, 2000), using case studies. Very few studies have taken a quantitative approach. The current work fills a gap in this respect. It takes a quantitative approach to analyze the effects of the possession of knowledge and access to external knowledge on the recognition of new business opportunities in Spain.

The rest of this work is organized as follows. Section 2 describes the eclectic theoretical framework chosen here to analyze the relations between knowledge and recognition of business opportunities. This section proposes the working hypotheses to be tested. The third section describes the methodological aspects of the research: the characteristics of the sample analyzed, the variable measurement and the statistical model used. Section 4 discusses the results obtained and their consequences for the verification of the proposed hypotheses. Finally, the last section discusses the most important conclusions of the research.

# Theoretical framework and hypotheses

An opportunity implies satisfying a market need through a creative combination of resources that provides superior value added (Schumpeter, 1934; Kirzner, 1973; Casson, 1982). The opportunity entails a market need that is defined vaguely, or alternatively a lack or misuse of certain resources or capabilities (Kirzner, 1973). Opportunities can originate in various ways, for example as a consequence of demographic, industrial or technological changes, or unexpected events, or incongruities in a particular market (Drucker, 1985). In this sense, an opportunity is a set of circumstances that creates a need or initiates a new business concept (Morris, 1998). Thus, opportunity recognition is considered the first stage in the entrepreneurial process (Timmons, 1990).

Opportunity recognition refers to the process by which someone perceives the possibility of creating a new and profitable business, product or service (Barringer and Ireland, 2007). Understanding how individuals discover and develop opportunities is a key part of entrepreneurship research (Venkataraman, 1997). What most of the entrepreneurship literature calls the opportunity recognition process consists of three distinct stages (Ardichvili et al., 2003): (1) perceiving market needs or underemployed resources; (2) discovering a fit between particular market needs and certain specified resources (Kirzner, 1973, 1979); and (3) creating the new fit between the hitherto separate needs and resources to create a new business concept (Hills, 1995; De Koning, 1999). Thus, the opportunity recognition process involves three activities (Miller, 2007): perception, discovery and creation.

Researchers have proposed a large number of models of opportunity recognition and development (Bhave, 1994; Schwartz and Teach, 1999; Singh et al., 1999a; De Koning, 1999; Sigrist, 1999). These models are based on different – often conflicting – assumptions borrowed from a number of disciplines ranging from cognitive psychology to Austrian School economics (Ardichvili et al., 2003).

Nevertheless, none of these models offers a complete explanation of the phenomenon. For example, Sigrist (1999) studies the cognitive process involved in opportunity recognition; De Koning (1999) and Hills et al. (1997) analyze the context of the social network; and Shane (2000) focuses on the prior knowledge and experience needed to perceive certain opportunities.

This theoretical gap suggests the need to develop eclectic theoretical models that complete the analysis of the identification of opportunities. In this respect, Ardichvili et al.'s (2003) model summarizes the main factors that influence the process by which individuals recognize a business opportunity, namely: (1) Kirzner's (1973) state of entrepreneurial alertness; (2) asymmetric information and prior knowledge (Hayek, 1945; Von Hippel, 1994; Shane, 2000); (3) social networks (Arenius and De Clercq, 2005; De Clercq and Arenius, 2006; Barringer and Ireland, 2007); (4) personality traits (e.g., optimism, perception of self-efficacy, creativity); and (5) the type of opportunity itself. As can be seen, these factors – apart from the last one – confirm that individuals differ in their capacity to recognize or perceive business opportunities (Shane, 2003). Whatever the type of opportunity, context or the way it originated, there is always a common factor: some individuals detect opportunities better than others (Shane, 2000, 2003).

Clarifying the elements that affect individuals' opportunity recognition requires reordering the idiosyncratic factors from Ardichvili et al.'s (2003) model. These elements can be regrouped into those that have something to do with the individual's intellectual capital (i.e., their knowledge, experiences and skills) and those that form part of the individual's social capital (i.e., their access to external knowledge offered by the social network).

In this line, the following subsections define the impact of intellectual capital and social capital on individuals' opportunity recognition, and offer a number of working hypotheses in this respect that are theoretically justified by the literature.

## *Intellectual capital*

Intellectual capital theorists consider that knowledge improves individuals' cognitive skills and allows them to work more productively and efficiently (Schultz, 1959; Becker, 1964; Mincer, 1974). Individuals with a higher quality intellectual capital should be better able to detect the existence of profitable business opportunities (Davidsson and Honig, 2003). The prior knowledge that comes from experience in work, education, or other sources influences the entrepreneur's capacity to understand, extrapolate, interpret and apply new information in a way that others cannot (Robert, 1991). Entrepreneurs discover opportunities because their prior knowledge triggers recognition of the value of the new information (Shane, 2000). The knowledge base that constitutes the intellectual capital and that could determine the individual's capacity to recognize business opportunities consists of, among other factors, their educational level, their knowledge and skills relating to business start-ups, and their previous experience as an entrepreneur.

With regard to the individual's educational level, Blanchflower (2004) finds that a positive relation exists between educational level and the creation of technology firms in rich countries. Indeed, emerging technologies are an important source of business opportunities (Thukral et al., 2008). According to Shane (2003), people with higher educational levels are more likely to start their own business and to be successful in doing so. The number of years individuals invest in their education is an important factor in successful start-ups, as is the acquisition of specific business management skills (Shane, 2003). If a link exists between educational level and firm creation, there should logically also be a relation with these individuals' capacity to detect business opportunities, since recognition is the first stage in firm creation. Arenius and De Clercq's (2005) recent work confirms this idea. These authors, in an empirical study in Belgium and Finland, find that educational level has a positive effect on individuals' probability of perceiving business opportunities. They explain this relation using two arguments. The first is that having a high educational level favours opportunity recognition by providing access to various types of knowledge. The second is that a large knowledge base increases the chance of relating this knowledge to potential opportunities (Cohen and Levinthal, 1990). Education can provide knowledge that is complementary to new, previously unavailable information potentially capable of generating a business opportunity. Consequently, having such a prior knowledge base may increase the ability to recognize opportunities (Shane and Venkataraman, 2000). Individuals have different knowledge bases, and so will differ in their ability to recognize the potential of certain opportunities. On the basis of the above arguments, the first hypothesis in this work is as follows:

> Hypothesis 1: Individuals with a university education are more likely to recognize good business opportunities in the area where they live than the rest of the population.

With regard to the second factor, which concerns the possession of knowledge and skills relating to business start-ups, Ajzen's (1991) theory of planned behaviour holds that if individuals consider they have the necessary skills, knowledge and ability to start their own business, they will have a positive and proactive attitude towards seeking business opportunities that allow them to do so. This attitude may increase the capacity to perceive opportunities. According to this theory, the stronger the belief that a particular, valuable objective (in this case, detecting an opportunity and exploiting it by starting a business) is feasible, the more likely the individuals will behave in such a way as to achieve that objective.

These skills and abilities are not necessarily linked to the educational level. In fact, some authors claim that entrepreneurs frequently possess a wide range of abilities without having an advanced or specific education (Murphy et al., 1991; Leazar, 2002).

Among the personality traits of entrepreneurs studied, two are associated with the successful recognition of opportunities: creativity and optimism (Ardichvili et al., 2003). Krueger and Dickson (1994) and Krueger and Brazeal (1994) show that entrepreneurial optimism is related to self-efficacy beliefs. Perceived self-efficacy makes the individual optimistic about their ability to achieve difficult specific objectives. The perception of self-efficacy, in other words, individuals' belief that they have the necessary abilities and knowledge (De Clercq and Arenius, 2006), guides them to optimism and towards a greater propensity to see opportunities rather than threats in any situation (Neck and Manz, 1992, 1996). An important literature supports this argument, and indicates that individuals could be more inclined to engage in activities relating to entrepreneurship (such as opportunity recognition) if they believe they have the skills needed to do these activities successfully (Scott and Twomey, 1988; Boyd and Vozikis, 1994; Chen et al., 1998; Davidsson and Honig, 2003; Arenius and Minniti, 2005; De Clercq and Arenius, 2006). The above arguments lead to the second hypothesis:

> Hypothesis 2: Individuals who are convinced they possess the knowledge and skills needed to start their own business are more likely to recognize good business opportunities in the area where they live than the rest of the population.

Finally, the third factor refers to the individual's previous experience as an entrepreneur. According to Shane (2000), entrepreneurs discover opportunities related to their prior knowledge. Individuals pay more attention to information that is related to the information they already have (Von Hippel, 1994). Entrepreneurship exists because of the information asymmetries between the different actors (Hayek, 1945). Conceivably, habitual entrepreneurs are particularly good at recognizing and developing opportunities (MacMillan, 1986; McGrath, 1996; McGrath and MacMillan, 2000). This is consistent with what McGrath (1996) and Ronstadt (1988) argue when they say that entrepreneurs who were previously entrepreneurs are better at detecting opportunities than others because of the specific knowledge generated by their previous entrepreneurial experiences. In the same line, other authors say that the knowledge accumulated as a result of previous entrepreneurial experiences may encourage opportunity generation (Ucbasaran et al., 2000; Ucbasaran and Westhead, 2002; Shane, 2003).

Various studies show that prior experience in a specific sector helps entrepreneurs to recognize business opportunities (Markham and Baron, 2003; Cooper and Park, 2008).

For example, an individual who works in a particular sector could identify a market niche in it that is not being served (Barringer and Ireland, 2007). Once the entrepreneur starts a firm and is immersed in a particular sector, new business opportunities begin to appear in the sector that are more difficult for someone from outside the sector to identify (Barringer and Ireland, 2007). An existing business can be a source of new ideas both in itself and through the experience the entrepreneur accumulates in that business, which provides the entrepreneur with information and skills that will be useful in the search for opportunities (Alsos and Kaikkonen, 2004). According to Shane (2000), individuals' prior knowledge consists of three dimensions that are important for opportunity discovery: prior knowledge of markets, prior knowledge of ways to serve those markets, and prior knowledge of customer problems. On the basis of the above, the third hypothesis is as follows:

> Hypothesis 3: Individuals who own and manage a firm are more likely to recognize good business opportunities in the area where they live than the rest of the population.

*Social capital*

Examination of the impact of the entrepreneur's knowledge base on their entrepreneurial activity does not end with the entrepreneur's own knowledge base. It is also necessary to consider the external knowledge that other people in the entrepreneur's environment provide and to which they are exposed, as the literature on networks and social capital describes (De Clercq and Arenius, 2006).

Social capital theory is closely related to network theory. Both theories consider individuals' capacity to extract benefits from the members of their social network (Lin et al., 1981; Portes, 1998). Social capital permits the social exchange (Emerson, 1972) of resources and information useful in the creation of businesses (Davidsson and Honig, 2003). The characteristics of an individual's social network affect opportunity recognition (Koller, 1988; Kingsley and Malecki, 2004). People who build a considerable network of social and professional contacts will be exposed to more ideas and opportunities (Cooper and Yin, 2005).

On the other hand, Singh et al. (1999b) argue that a large social network with numerous weak ties with people outside the circle of close friends and relatives is positively associated with the identification of ideas and the recognition of opportunities. Entrepreneurs are more likely to obtain a business idea through a weak tie, with a casual acquaintance with whom they have little interaction, than through a strong tie, with a colleague, friend or relative with whom they have frequent interaction (Singh et al., 1999b) and who has a similar cognitive map to themselves.

But undoubtedly the most valuable social network in the field of entrepreneurship is one that contains other entrepreneurs and includes, in turn, their social networks. Knowing other entrepreneurs, or even being a business angel in a business, favours individuals' opportunity recognition capacity. In this line, the impact of social capital on the recognition of business opportunities comes from, among other factors, knowing other entrepreneurs and being a business angel.

The first factor refers to whether the individual knows other entrepreneurs. Various empirical studies stress the importance of indirect experience in favouring the propensity to create a firm (Scherer et al., 1991; Delmar and Gunnarsson, 2000). According to role theory (Veciana, 1999), people who know other entrepreneurs, either from their close geographic environment or through more or less direct relationships (friends, relatives, etc.), may capture facts from them that make the idea of creating a firm and being successful in doing so more credible. Thus, individuals who can capture and replicate "entrepreneur roles" will be more likely to be attentive to opportunities allowing them to become an entrepreneur themselves.

From the network theory perspective, social networks can provide critical information, ideas and resources for starting a new firm (Larson and Starr, 1993). According to Arenius and De Clercq (2005), some people are more exposed to receiving new information, and consequently recognizing business opportunities, as a result of the social networks to which they belong. In this line, Hills et al. (1997) argue that entrepreneurs with large networks identify a significantly larger number of opportunities than entrepreneurs lacking such networks (solo entrepreneurs). According to these authors, the quality of the network of contacts can affect the opportunity identification process. If a member of the potential entrepreneur's social network is already an entrepreneur, the quality and utility of the ideas, information and resources that they can provide will be even greater (De Clercq and Arenius, 2006).

It is also important to consider the weak ties (Granovetter, 1973) between the potential entrepreneur and current entrepreneurs, as well as those between the potential entrepreneur and people belonging to the current entrepreneurs' social networks. Weak ties have greater value added because people with mental patterns and experiences that are very different from those of the potential entrepreneur will offer very different perspectives. Weak ties give greater access to unique information than strong ties (Arenius and De Clercq, 2005). Moreover, people can only maintain a limited number of strong ties with others, while they can have a much higher number of weak ties (Granovetter, 1973).

According to Alsos and Kaikkonen (2004), an existing business can be a continuous source of new ideas through the social networks developed by the owner and the management. In this line, McGrath (1996) argues that entrepreneurs with access to a large, well-functioning network – for example through an existing business – could identify a large number of high quality latent business opportunities. In turn, McAdam and Marlow (2007) point out that places where entrepreneurs meet, such as business incubators, offer access to established business networks and to opportunities. On the basis of the above, the fourth hypothesis is as follows:

> Hypothesis 4: Individuals who know another entrepreneur are more likely to recognize good business opportunities in the area where they live than the rest of the population.

The second factor proposed is whether the potential entrepreneur is a business angel. Business angels are a specific case of the previous one, in other words, their investment in other people's businesses means they cannot avoid knowing one or more entrepreneurs. Being a business angel could imply a closer relationship than simply knowing an entrepreneur but not participating in their business in any way.

The theoretical arguments offered above for individuals who know entrepreneurs are even stronger here. Replicating the roles of known entrepreneurs (Veciana, 1999) or exchanging information and ideas with them (Larson and Starr, 1993) increases business angels' capacity to perceive new opportunities in which to invest, either by informal investment or by starting a business. According to De Clercq and Arenius (2006), the experience of being a business angel conceivably entails exposure to other entrepreneurs' knowledge, even if the investment involves a purely financial, not personal, relationship. These considerations lead to the final hypothesis of this work:

> Hypothesis 5: Individuals who have provided personal funds to help others start a business are more likely to recognize good business opportunities in the area where they live than the rest of the population.

# Methodology

**Sample**

In order to test the hypotheses the authors use the Adult Population Survey (APS) from the Global Entrepreneurship Monitor (GEM) project. Telephone interviews were conducted during April-June 2007 with the 27,880 respondents using a standardized questionnaire designed by the GEM Consortium's research team and translated from English into the native language of the country (Spanish). The interviews were undertaken by Opinometre Institut, an independent market research firm. The sampling error was ±0.58% with a 95% confidence level.

The respondents formed part of the Spanish adult population between 18 and 64 at the time of the interview. For a more detailed explanation of the survey process, see Reynolds et al. (2005).

**Measures**

The variables used in this work are measured as follows.

Dependent variable

*Opportunity recognition*. The respondents were asked if "in the next six months there will be good opportunities for starting a business in the area where [they] live". This is a binary variable that equals 1 if the answer is affirmative and 0 otherwise.

Predictor variables

The predictor variables have been grouped according to their relation to the existing knowledge base or to exposure to external knowledge.

a. Existing knowledge base (intellectual capital):

*Educational level.* The aim is to understand the role of university education in the level of individuals' opportunity recognition, so the responses are coded 1=university education and 0=other studies (including no education).

*Knowledge and skills required to start a business.* To measure this variable the work uses the individuals' perception of their capacity to start a business. Specifically, they are asked if they "have the knowledge, skills and experience required to start a new business". The responses are coded as 1=yes and 0 otherwise.

*Owning and managing a firm*: To detect if the individual owns and manages a firm (at the time of the interview), they are asked if they "are, alone or with others, currently the owner of a company [they] help manage, self-employed, or selling any goods or services to others". This dichotomous variable equals 1 if the respondent responds affirmatively and 0 otherwise.

b. Access to external knowledge (social capital):

*Contacts with entrepreneurs*: To find out if the individual has an entrepreneur among their personal relationships, they are asked if they "know someone personally who started a business in the past two years". This dichotomous variable equals 1 if the respondent responds affirmatively and 0 otherwise.

*Business angel*: To determine if the individual has this type of contact with entrepreneurs, they are asked if they "have, in the past three years, personally provided funds for a new business started by someone else, excluding any purchases of stocks or mutual funds". This dichotomous variable equals 1 if the respondent responds affirmatively and 0 otherwise.

Control variables

Apart from the above variables the estimated model also includes *age*, *gender* and *habitat* (residence in a rural or urban area) as control variables. *Age* is a continuous variable taking values between 18 and 64 years. *Gender* is a dichotomous variable with values 0=male and 1=female. *Habitat* is also a dichotomous variable, and takes values 1=rural (the respondent's town of residence has a population of less than 10,000) and 0=urban.

**Econometric model**

*Specification*: the research question here is to evaluate the influence of a series of independent, or explanatory, variables on a dichotomous dependent variable, so the appropriate econometric model is the general logistic regression model. The authors follow Hosmer and Lemeshow's (2000) recommendations about how to estimate the model.

The model specification corresponds to the following expression:

$$P(Y=1) = \frac{1}{1+e^{-(\beta'X)}}$$

The vector X consists of the explanatory variables presented above; β is the vector of unknown model parameters that measure the impact of each variable on the probability that the endogenous variable (i.e., opportunity recognition) equals 1 given the observed values in X.

*Estimation*. The vector of unknown parameters of the model is estimated from the sample information using the maximum likelihood function as loss function. The authors use the statistics package SPSS (version 14.1) for this. To evaluate the possible existence of multicollinearity between the model variables the authors inspect the values in the correlation matrix of the independent variables.

*Verification*. Model verification is through the likelihood ratio test. The authors also use the Hosmer-Lemeshow test and the percentage of correct classifications of the observed values compared to the values predicted by the model. The Wald test is used to test the statistical significance of the regression coefficients.

The control variables are age, gender and habitat. These variables are not of primary interest in this research, but they do also affect the dependent variable, so they are included to control for their effects.

Apart from the above, and in order to facilitate interpretation of the results the authors calculate the "odds ratio" (the exponent of the regression coefficient). In dichotomous variables the odds ratio represents how much more probable it is for the phenomenon of interest (here, opportunity recognition) to occur in individuals who score 1 on the predictor variable compared to those who score 0. For example, for the variable gender, which is coded 0=male and 1=female, an odds ratio of 2 would mean women are twice as likely to recognize business opportunities as men.

*Exploitation*. The exploitation of the adjusted model consists of verifying the statistical significance of the parameters, identifying their sign and discussing the implications, and testing the hypotheses in order to evaluate the role of the predictor variables in the opportunity recognition phenomenon.

# Results

Table 1 shows the descriptive statistics and the correlation matrix for the variables analyzed. As the table shows, 24% of the individuals analyzed say they can see good business opportunities in the area where they live. A slightly higher percentage has a university education (25%), while 46% claim to have the knowledge and skills required to start a business. Meanwhile, 12% say they own and manage a firm, 3% have been a business angel, and 35% say they have personal contacts with entrepreneurs. Among the control variables, the average age of the respondents is 41.65, half are men and half are women, and only a small minority lives in rural areas (16%).

On the other hand, the correlations between the predictor variables are not high, which indicates a low probability of multicollinearity between the variables and a consequent presence of undesired effects on the estimations of the parameters. The table also shows that the probability individuals see good business opportunities is positively correlated with all the variables examined in the working hypotheses, which provides preliminary support for them.

*[Please insert table 1 about here]*

The likelihood ratio test is used to test the null hypothesis that all the model coefficients equal 0, compared to the existence of at least one non-zero parameter. The results show that the null hypothesis can be rejected at the 1% level, which means the goodness of fit of the model is satisfactory ($\chi^2$=1270.503; p<0.01).

On the other hand, and as a complement to the previous test, the Hosmer-Lemeshow test tests the null hypothesis that the model fits the data adequately. If this hypothesis is true, the test statistic should follow a $\chi^2$ distribution with as many degrees of freedom as independent variables in the model. If the statistical significance of the test is low (i.e., less than 0.05) the model does not fit the data adequately. The results here give a Hosmer-Lemeshow statistic of 0.181, so the model can be considered satisfactory.

Finally, the adjusted model has a correct classification rate of 75.6%. In other words, given the observed values of the predictor variables the model correctly predicts the observed value of the dependent variable in 3 out of 4 individuals.

Table 2 shows the results of the estimation of the model. As the significance values of the Wald statistic show, all the variables are statistically different from zero at the 5% level except for owning and managing a firm (p=0.202 > 0.05). Thus, all the variables except owning and managing a firm affect the probability that the phenomenon of interest here (recognition of good business opportunities) will occur.

The results for the control variables show that both age ($\beta$=-0.007; p<0.01) and habitat ($\beta$=-0.090; p<0.01) have a weak negative effect on the dependent variable. In other words, the older the potential entrepreneur, the lower their capacity to recognize business opportunities, and it is more difficult to find such opportunities in rural areas than in urban areas. On the other hand, and as in Arenius and De Clercq (2005), the results show a difference in the capacity to recognise good business opportunities between the genders in favour of men ($\beta$=-0.197; p<0.01).

*[Please insert table 2 about here]*

Table 3 summarizes the results of the hypothesis tests, with the hypotheses grouped according to their relation to the individual's possession of knowledge (intellectual capital) or access to external knowledge (social capital).

*[Please insert table 3 about here]*

The results obtained demonstrate the impact on capacity to recognize business opportunities of the following factors: higher education, the perceived possession of the skills and knowledge required to start a business, being a business angel, and knowing other entrepreneurs.

The analysis carried out here provides support for four of the five hypotheses proposed, and offers an approximation to the factors that favour greater capacity to recognize business opportunities in individuals. The size of the sample analyzed and the significance level of the model support the validity of the results obtained.

According to these results, knowing other entrepreneurs is the most important factor in the development of the ability to detect opportunities for new firm creation. The presence of this factor almost doubles individuals' capacity to recognize business opportunities. These contacts clearly serve as points of reference and models to the potential entrepreneurs.

The next most influential factors, which have a similar explanatory power, are being a business angel and the perceived possession of the knowledge and skills required to start a business. Possessing either of these characteristics makes it over 50% more likely individuals will recognize business opportunities.

University education has a positive effect on the capacity to recognize business opportunities, although according to these results this factor is only weakly significant, and the difference between individuals with a university education and those with a lower educational level is slight.

# Conclusions

Recognizing business opportunities is not easy. The opportunities are there, but only some individuals have the skills required to recognize them, and since this stage is fundamental for the success of the entrepreneurial process, it is worth studying the determining factors of these skills. This work has aimed to identify some of the variables relating to the individual's intellectual and social capital that influence the process of recognizing these opportunities. After a review of the literature, the authors formulated a number of hypotheses, which they tested using the sample from the GEM Consortium's survey of the Spanish adult population in 2007. The size of the sample and the previous validation of the questionnaires used reinforce the reliability of the results obtained here (Reynolds et al., 2005).

The empirical study carried out here supports the view that both intellectual capital and social capital have a positive effect on people's capacity to recognize business opportunities. This effect can be explained by the presence of factors such as contact with other entrepreneurs, having supported other entrepreneurs' projects through business angel funding, the personal conviction of having the knowledge and skills needed to start a business, and university education. The first two factors are part of the individual's social capital and the other two are part of their intellectual capital. The results obtained here confirm the positive impact of both forms of capital on the ability to recognize new business opportunities, although the explanatory power of the social capital variables seems to be greater.

These results are consistent with and complementary to those of other recent studies (Ozgen and Baron, 2007; De Carolis et al., 2009) that find a relation between the entrepreneur's social capital and opportunity recognition and the chance of success in firm creation.

Thus, the difference between individuals capable of recognizing good business opportunities and those not capable is due more to the individuals' relationships with other entrepreneurs than to their personal knowledge and experience. Although the intellectual capital factors do have a positive impact, individuals' access to external knowledge through the social networks in which they participate proves to be fundamental for developing the capacity to recognize new business opportunities.

Unlike the results of some previous research (Barringer and Ireland, 2007), the current work has not found empirical evidence that having entrepreneurial experience ("owning and managing a firm" variable) has a positive effect on capacity to recognize opportunities. This result could be due to the question asked in the questionnaire to measure this variable. This variable can only distinguish between individuals who own and manage a firm and those who are in a different situation in the labour market (e.g., working for an employer) and who may also have some type of previous experience in the sector in which they work. This would explain the result, in which being a current entrepreneur does not imply a greater propensity to recognize opportunities than the rest of the population.

As can be deduced from the results of this study, the information asymmetry that allows some individuals to recognize opportunities but not others comes mainly from their access to social rather than intellectual capital, although the latter type of capital also plays an important role. But up to now government policies designed to encourage entrepreneurial activity have mainly focused on the individual, in other words, they have mainly centred on developing personal capabilities. Encouragement of relations and contact between current and potential entrepreneurs has been considered far less important.

These results lead to an important implication for the public authorities. Entrepreneurs' social networks generate important benefits for the population as a whole in terms of recognition of business opportunities, so public authorities wishing to promote entrepreneurship must offer more support for the creation, expansion and diffusion of these social networks. Personal contact, the exchange of experiences, and the diffusion of the values of success based on the entrepreneur's effort and perseverance encourage a demonstration effect based on the example of people who have created their own businesses.

Thus, encouraging entrepreneurship in the population of a particular country or area requires the creation and strengthening of social networks that facilitate links between potential entrepreneurs and current entrepreneurs. The transmission of knowledge coming from other entrepreneurs' experiences emerges as an important factor for increasing the level of entrepreneurship in the population.

It should be borne in mind that intellectual capital is still important for opportunity recognition. Thus, the authors can offer another recommendation for the design of government programmes for the promotion of entrepreneurship in a country. Superior value added can be obtained by encouraging social networks that include the population of entrepreneurs with the greatest intellectual capital. In other words, it would be particularly useful to encourage the creation of social networks between entrepreneurs with a higher educational level and experience. Specifically, the authors would recommend that the public authorities develop social networks of entrepreneurs in the universities and research centres. Social networks that include university-educated people, members of R&D centres, or people who have participated in the creation of university spin-offs will have significant potential for opportunity recognition.

Boyd, N. G; Vozikis, G. S. (1994) "The Influences of Self-efficacy on the Development of Entrepreneurial Intentions and Actions." *Entrepreneurial Theory and Practice* 18(4) 63–90.

Casson, M. (1982) *The Entrepreneur*. Barnes and Noble Books, Totowa, NJ.

Chen, C.C.; Greene, P.G.; Crick, A. (1998) "Does entrepreneurial self-efficacy distinguish entrepreneurs from managers?" *Journal of Business Venturing* 13(4) 295-316.

Cohen, W.M.; Levinthal, D.A. (1990) "Absorptive Capacity: A New Perspective on Learning and Innovation." *Administrative Science Quarterly* 35: 128–152.

Cooper, A.C.; Yin, X. (2005) "Entrepreneurial Networks." in *The Blackwell Encyclopedia o/ Management-Entrepreneurship*, eds. M. A. Hitt and R. D. Ireland (Malden, MA: Blackwell Publishing) 98-100.

Cooper, S.Y.; Park, J.S. (2008) "The impact of 'incubator' organizations on opportunity recognition and technology innovation in new, entrepreneurial high-technology ventures." *International Small Business Journal* 26(1) 27-56.

Corbett, A.C. (2002) "Recognizing High-Tech Opportunities: A Learning and Cognitive Approach." In *Frontiers of Entrepreneurship Research*. Wellesley, Mass.: Babson College.

Craig, J.; Lindsay, N. (2001) "Quantifying 'Gut Feeling' in the Opportunity Recognition Process." In *Frontiers of Entrepreneurship Research*. Wellesley, Mass.: Babson College.

Davidsson, P.; Honig, B. (2003) "The role of social and human capital among nascent entrepreneurs". *Journal of Business Venturing* 18: 301-331.
25

**Table 1. Descriptive statistics and correlation matrix**

|  | Mean | SD | 1 | 2 | 3 | 4 | 5 | 6 | 7 | 8 |
|---|---|---|---|---|---|---|---|---|---|---|
| 1. Perceives good opportunities in next 6 months | .24 | .430 |  |  |  |  |  |  |  |  |
| 2. University education | .25 | .431 | .050(**) |  |  |  |  |  |  |  |
| 3. Has knowledge and skills to start business | .46 | .498 | .136(**) | .103(**) |  |  |  |  |  |  |
| 4. Owns and manages firm | .12 | .320 | .053(**) | .019(**) | .287(**) |  |  |  |  |  |
| 5. Has been business angel in past 3 years | .03 | .174 | .064(**) | .047(**) | .072(**) | .050(**) |  |  |  |  |
| 6. Has known entrepreneur personally in past 2 years | .35 | .476 | .177(**) | .089(**) | .234(**) | .114(**) | .115(**) |  |  |  |
| 7. Age | 41.65 | 12.448 | -.064(**) | -.070(**) | -.029(**) | .042(**) | .004 | -.159(**) |  |  |
| 8. Gender | .50 | .500 | -.068(**) | -.020(**) | -.104(**) | -.060(**) | -.027(**) | -.106(**) | .052(**) |  |
| 9. Habitat | .16 | .366 | -.014(*) | -.049(**) | .015(*) | .056(**) | .009 | -.009 | .039(**) | -.025(**) |

** Correlation significant at 0.01 level (two-tail).

* Correlation significant at 0.05 level (two-tail).

**Table 2. Logistic regression. Dependent variable "Opportunity recognition"**

|  |  | β | SD | Wald | df | Sig. | Exp(β) |
|---|---|---|---|---|---|---|---|
| Step 1(a) | education | .120 | .032 | 13.718 | 1 | .000 | 1.128 |
|  | skills | .443 | .031 | 210.830 | 1 | .000 | 1.557 |
|  | own and manage | .056 | .044 | 1.625 | 1 | .202 | 1.058 |
|  | bus. angel | .437 | .073 | 35.824 | 1 | .000 | 1.548 |
|  | know | .652 | .030 | 471.266 | 1 | .000 | 1.920 |
|  | age | -.007 | .001 | 37.563 | 1 | .000 | .993 |
|  | gender | -.197 | .029 | 46.585 | 1 | .000 | .821 |
|  | habitat | -.090 | .040 | 5.157 | 1 | .023 | .914 |
|  | Constant | -1.253 | .057 | 483.325 | 1 | .000 | .286 |

(a) Variable(s) introduced in step 1: education, skills, own and manage, bus. angel, know, age, gender, habitat.

**Table 3. Summary of results**

|  | **Hypothesis** | **Variable** | **Impact** |
|---|---|---|---|
| **Intellectual capital** | **Hypothesis 1.** Individuals with a university education are more likely to recognize good business opportunities in the area where they live than the rest of the population. | Educational level | Positive |
| | **Hypothesis 2.** Individuals who are convinced they possess the knowledge and skills needed to start their own business are more likely to recognize good business opportunities in the area where they live than the rest of the population. | Possession of knowledge & skills | Positive |
| | **Hypothesis 3.** Individuals who own and manage a firm are more likely to recognize good business opportunities in the area where they live than the rest of the population. | Own and manage firm | Not significant |
| **Social capital** | **Hypothesis 4.** Individuals who know another entrepreneur are more likely to recognize good business opportunities in the area where they live than the rest of the population. | Contact with entrepreneurs | Positive |
| | **Hypothesis 5.** Individuals who have provided personal funds to help others start a business are more likely to recognize good business opportunities in the area where they live than the rest of the population. | Experience as business angel | Positive |